\documentclass[11pt,dvips]{article}

\usepackage{epsfig,times} 
\usepackage{picinpar}
\usepackage{wrapfig}
\usepackage{floatflt}
\makeatletter
\renewcommand{\section}{\@startsection{section}{1}{0in}
	{0.4\baselineskip}{0.1\baselineskip}{\Large\bf}}
\renewcommand{\subsection}{\@startsection{subsection}{2}{0in}
	{0.25\baselineskip}{-\baselineskip}{\large\bf}}
\renewcommand{\subsubsection}{\@startsection{subsubsection}{3}{0in}
	{0.1\baselineskip}{-\baselineskip}{\normalsize\bf}}
\makeatother
\begin{document}
%
\thispagestyle{myheadings}
%
\markright{HE 4.1.06}
\begin{center}
%
{\LARGE \bf Low Energy Atmospheric Neutrino Events in MACRO}
\end{center}

\begin{center}
%
%
{\bf A. Surdo$^1$ for the MACRO Collaboration}\\
{\it $^1$ INFN Sezione di Lecce - Via Arnesano, I-73100 Lecce, Italy}
\end{center}

\begin{center}
{\large \bf Abstract\\}
\end{center}
\vspace{-0.5ex}
%
%
The flux of low energy atmospheric neutrinos 
($\overline E_\nu \sim 4\ GeV$)
has been studied with the MACRO detector at Gran Sasso by detecting
$\nu_\mu$ interactions inside the apparatus, and by upward-going stopping 
muons. The updated analysis of the data collected until now with the complete
apparatus will be presented. The results show a deficit of the
measured number of events in an uniform way over the whole zenith angle with
respect to Monte Carlo predictions. The deficit and the angular
distributions, when interpreted in terms of neutrino oscillations, are
consistent with the MACRO results on the much higher energy upward
throughgoing muons ($\overline E_\nu \sim 100\ GeV$).
%

\vspace{1ex}

%
%
\section{Introduction:}
\label{intro.sec}

Recent results (Fukuda, 1998, Ambrosio, 1998a) have confirmed the
anomaly in the atmospheric neutrino flux which was previously observed 
by several underground experiments (Casper, 1991, Fukuda, 1994, 
Allison, 1997). The suggested explanation for this anomaly is $\nu_\mu$ 
disappearance due to neutrino oscillations, with maximum mixing and 
$\Delta m^2$ in the range of a few times 10$^{-3}$ eV$^2$. The high 
energy $\nu_\mu$ events have been deeply investigated by the MACRO 
experiment (Ronga, 1999). Here we 
report on the updated analysis (Bernardini, 1998, Spurio, 1998) of low 
energy $\nu$ events.


The MACRO detector (Ahlen, 1993)
is a large rectangular box (76.6~m~$\times$~12~m~$\times$~9.3~m)
whose active detection elements are planes of streamer tubes for tracking
and liquid scintillation counters for fast timing.
 The lower half of the detector is filled
with trays of crushed rock absorber alternating with streamer tube
planes, while the upper part is open.
The low energy $\nu_\mu$ flux can be studied by the detection of $\nu_\mu$ 
interactions inside the apparatus, and by the detection of upward going 
muons produced in the rock surrounding it and stopping inside the detector 
(Fig. 1a). Because of the MACRO geometry, muons induced by neutrinos with 
the interaction vertex inside the apparatus can be tagged with 
{\it time-of-flight} ($T.o.F.$) measurement only for upgoing muons 
({\it $IU\mu$=Internal Upgoing $\mu$}). The downgoing muons with vertex in 
MACRO ({\it $ID\mu$=Internal Downgoing $\mu$}) and upward going muons
stopping inside the detector ({\it $UGS\mu$ = Upward Going Stopping $\mu$})
can be identified via topological constraints.
Fig. 1b shows the parent neutrino energy distribution for the three event
topologies detected by MACRO.
The {\it Internal Upgoing $\mu$} events are produced by 
parent neutrinos with energy spectrum almost equal to that of the 
{\it Internal Downgoing plus Upward Going Stopping $\mu$} events.

\vskip 0.4 cm
\section{\bf Internal Upgoing Events ($IU$):}
\label{iup.sec}
%
%
%
%
The data sample used for the Internal Upgoing ($IU$) events corresponds to an
effective live-time of 4.1 years from April, 1994 up to February, 1999.
The identification of $IU$ events was based both on topological
criteria and $T.o.F.$ measurements.
The basic requirement is the presence of at least two
scintillator clusters in the upper part of the apparatus (see Fig. 1a)
matching a streamer tube track reconstructed in space. A similar
request is made in the analysis for the up throughgoing events produced by $\nu_
\mu$ interactions in the rock below the detector (Ambrosio, 1998a).

For $IU$ candidates, the track starting point must be inside the
apparatus. To reject fake semi-contained events entering from a
detector crack, the extrapolation of the track in the lower part
of the detector must cross and not fire at least three streamer tube planes
and one scintillation counter. 

\begin{center}
\begin{figure}
\hspace{4ex}
\epsfig{figure=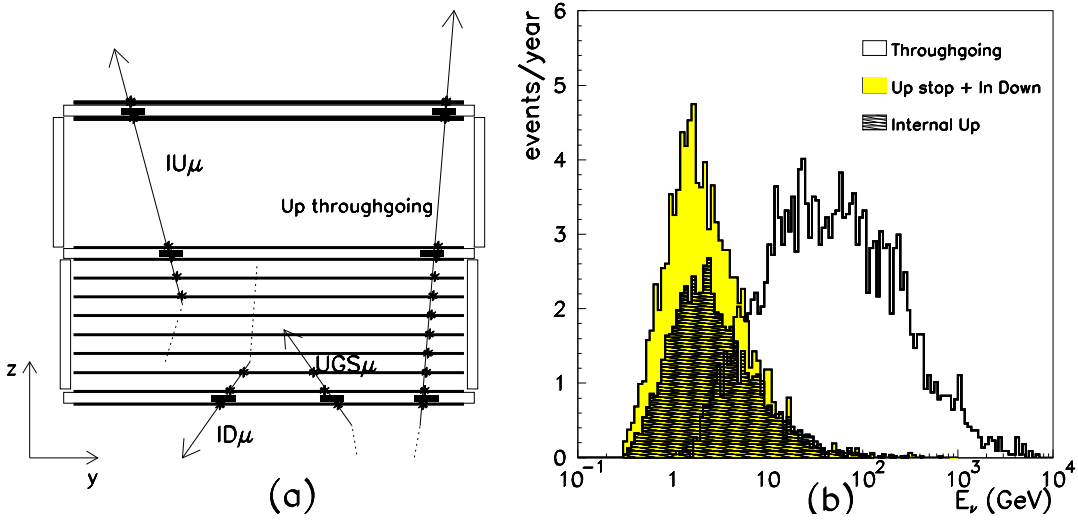,height=7cm}
{\caption {\label{fig:topo}\small Left: sketch of
different event topologies induced by neutrino interaction in or around MACRO.
$IU\mu$= Internal Upgoing $\mu$;
$ID\mu$= Internal Downgoing $\mu$;
$UGS\mu$= Upgoing Stopping $\mu$;
Up throughgoing = upward throughgoing $\mu$.
In the figure, the stars represent the streamer tube hits, and the black 
boxes the scintillator hits. The $T.o.F.$ of the particle can be measured 
for the $IU\mu$ and up throughgoing events. Right: parent neutrino energy
distribution for the three $\nu_\mu$ samples.
}}
\end{figure}
\end{center}

The above conditions, tuned on the
Monte Carlo simulated events,
account for detector inefficiencies and reduce
the contribution from upward throughgoing muons which mimic
semi-contained muons to less than $\sim 1 \%$.
The measured $1/\beta$ distribution is shown in Fig.~2.
The measured muon velocity $\beta c$ is calculated with the
convention that downgoing muons have
1/$\beta$ near +1 while upgoing muons have 1/$\beta$ near -1.
It was evaluated that $5$ events are due to an uncorrelated background.
After the background subtraction, $116$ events are classified as $IU$ events.


\vskip 0.4 cm
\section{\bf Upgoing Stopping Events ($UGS$) and Internal Downgoing ($ID$):}
\label{iud.sec}

The $UGS + ID$ events are identified via topological constraints, and not
with the $T.o.F$. For this analysis, the effective live-time is $4.1\ y$.
The main request for the event selection is
the presence of one reconstructed track crossing
the bottom layer of the scintillation counters (see Fig. 1a).
All the hits along the track must be confined at least one meter inside
each wall of a MACRO supermodule.
The selection conditions for the event vertex (or $\mu$ stop point)
in the detector are 
symmetrical to those for the $IU$ search, and reduce to a negligible level
the probability that an atmospheric muon produces a background event.
The main difference with respect to the $IU$ analysis (apart from the 
$T.o.F.$) is that on average fewer streamer tube hits are fired.
To reject ambiguous and/or wrongly tracked events
which passed the event selection, a scan with the
MACRO Event Display was performed.
All the real and simulated events which passed the event selection
were randomly merged. The accepted events passed a double scan procedure
(differences are included in the systematic uncertainty).

The main background source is due to upward going charged particles
(mainly pions) induced by interactions of atmospheric muons in the rock
around the detector (Ambrosio, 1998b).

\vskip 0.4 cm
\section{\bf Comparisons between Data and Monte Carlo:}

\begin{figwindow}[1,r,%
{\mbox{\epsfig{file=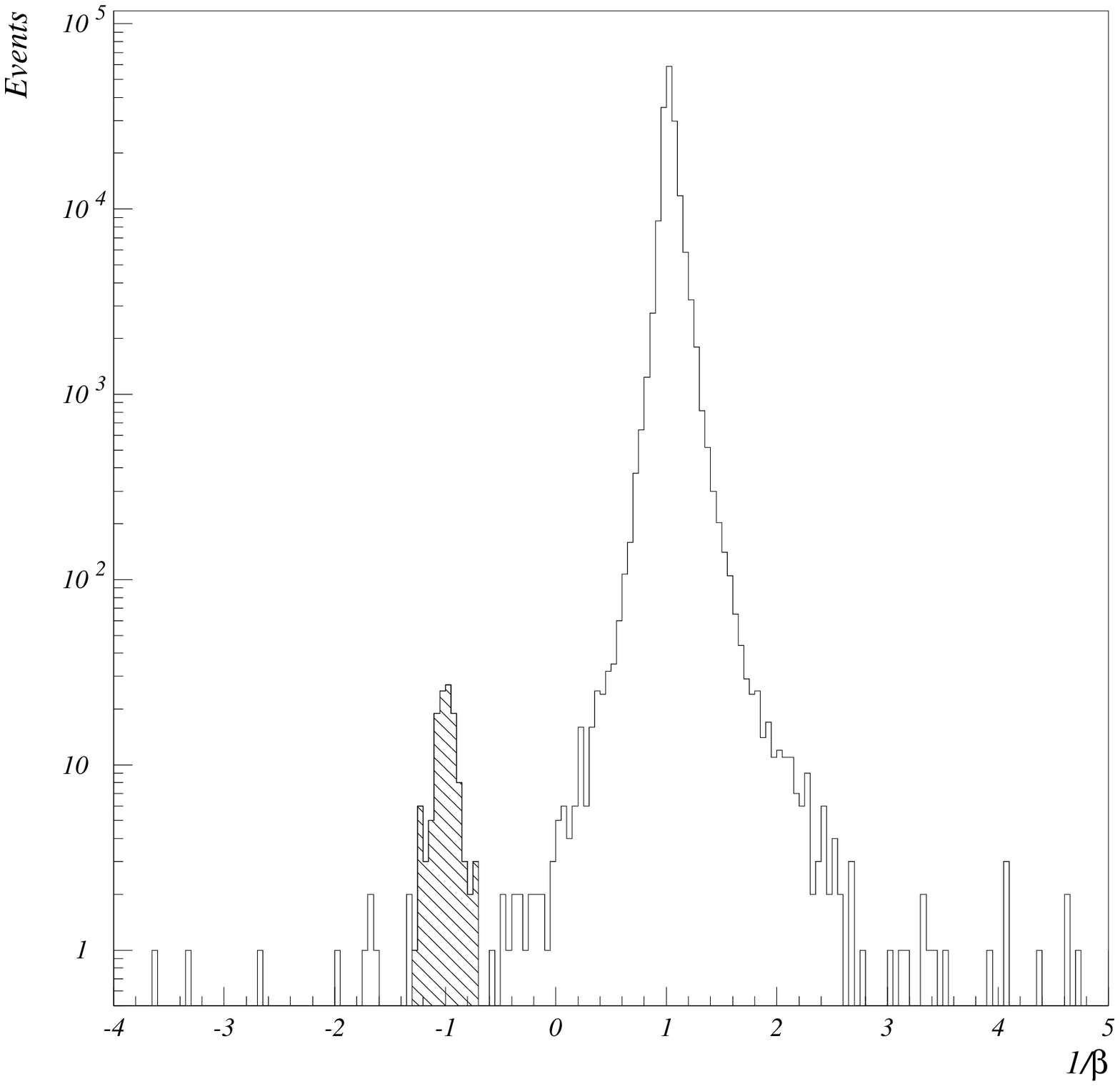,width=3.0in}}},%
{The $1/\beta$ distribution of the detected $IU$ events (dashed area) after 
all analysis cuts. The remaining $\sim 1.6\times 10^5$ are downgoing 
atmospheric stopping muons.}] 
%
%
%
The expected rates were evaluated with a full Monte Carlo
simulation. The events are mainly due to $\nu_\mu$ CC, with a
contribution from NC and $\nu_e$ ($\sim 13 \%$ for $IU$ and $\sim
10 \%$ for $UGS + ID$). The $\nu_e$ and $\nu_\mu$ were allowed to
interact in a volume of rock containing the experimental Hall B
and the detector.  The rock mass in the generation volume is
$169.6\ kton$, while the MACRO mass is $5.3\ kton$. The
atmospheric $\nu$ flux of the Bartol group (Agrawal, 1996) and the
cross sections of Lipari 1995 were used. The detector response has
been simulated using GEANT and simulated events are processed in
the same analysis chain as the real data. In the simulation, the
parameters of the streamer tube and scintillator systems have been
chosen in order to reproduce the real average efficiencies. The
total theoretical uncertainty on $\nu$ flux  and  cross section at
these energies is of the order of $25\%$. The systematic error  is
of the order of $10\%$, arising from the simulation of detector
response, data taking  conditions, analysis algorithm efficiency,
and the mass and acceptance of the detector. With our full MC
simulation, the prediction for $IU$ events is $202 \pm 20_{syst}
\pm 50_{theor}$, while the observed number of events is $116 \pm
11_{stat}$. The ratio $R=(DATA / MC)_{IU} =0.57 \pm 0.05_{stat}
\pm 0.06_{syst} \pm 0.14_{theor}$. 
The prediction for $UGS + ID$ events is $274 \pm 27_{syst} \pm
68_{theor}$, while the observed number of events is $193 \pm
14_{stat}$. The ratio $R=(DATA / MC)_{UGS + ID}=0.71 \pm
0.05_{stat} \pm 0.07_{syst} \pm 0.18_{theor}$. An almost equal
number of $UGS$ and $ID$ neutrino induced events are expected in
our data sample. Fig. 3 shows the angular distribution of the $IU$
and $UGS + ID$ data samples, with the Monte Carlo predictions. 
\end{figwindow}

\vskip -1.0 cm

\begin{figure}[bh!]
\epsfig{file=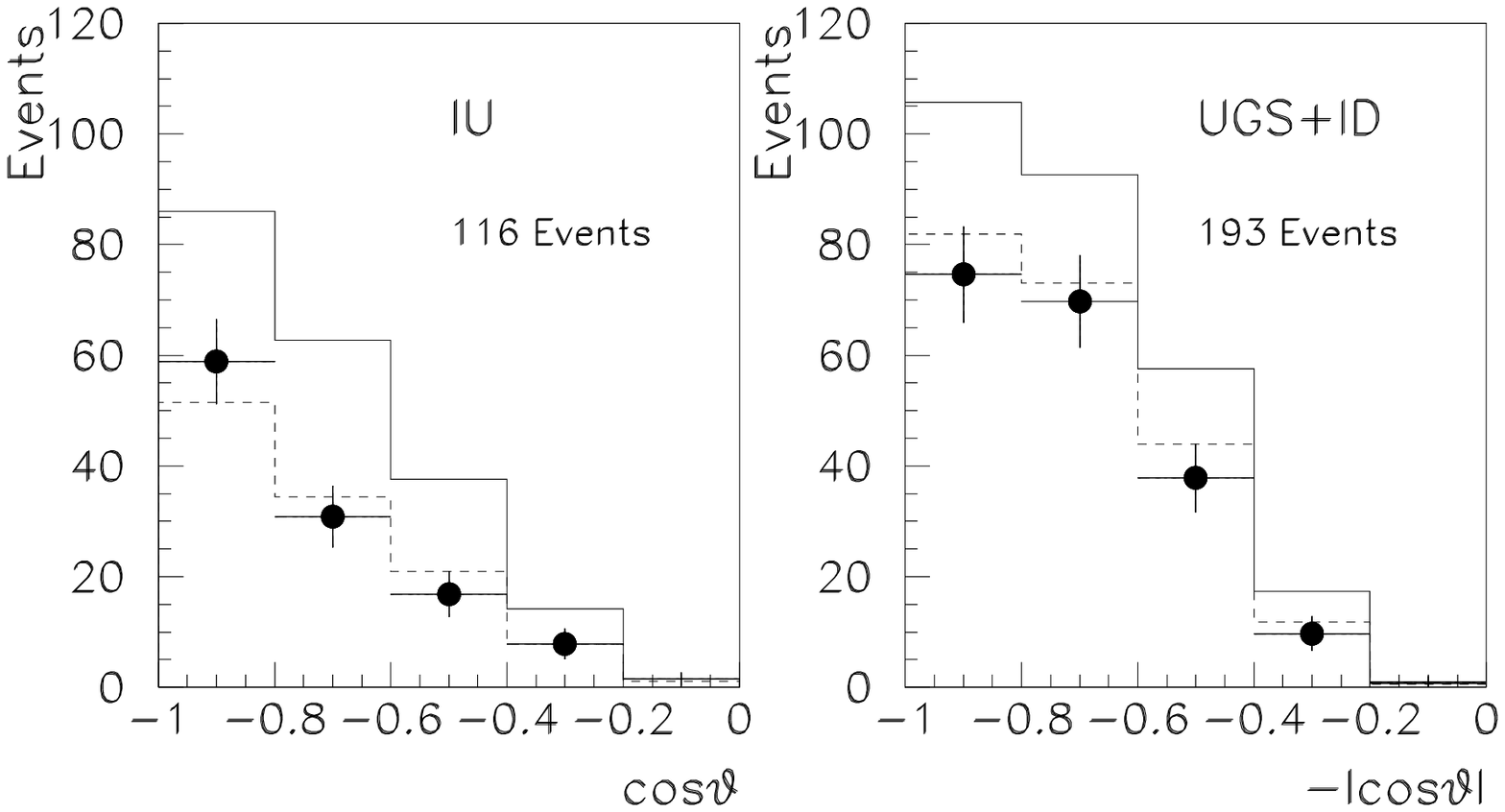,height=10.5cm,width=16.5cm}
\vskip -0.9 cm
\caption{\label{fig:inboler}\small
Zenith angle ($\theta$) distribution for $IU$ and $UGS + ID$ events.
The background-corrected data points (black points with error bars)
are compared with the Monte Carlo
expectation assuming no oscillation (full line) and two-flavour
oscillation (dashed line) using maximum mixing and 
$\Delta m^2=2.5 \times 10^{-3}\ eV^2$.}
\end{figure}

The low energy $\nu_\mu$ samples show a deficit of the measured number of
events  over the whole angular distribution
with respect to the predictions based on the absence
of neutrino oscillations.
The measured deficit of low-energy events is in agreement with
the MACRO results on the throughgoing events (Ambrosio, 1998a, Ronga, 1999),
{\it i.e.} with a model of $\nu_\mu$ disappearence
with $\sin^2 2\theta \simeq 1.0$ and
$\Delta m^2 \sim 2.5 \times 10^{-3}\ eV^2$.
In fact, the $IU$ and $UGS$ events have crossed the Earth 
($L\sim 13000\ km$), and in the
energy range of few $GeV$ the flux is reduced by a factor of two for 
maximum
mixing and $\Delta m^2 \sim 10^{-2} \div 10^{-3}\ eV^2$. No flux reduction
is expected for $ID$ events ($L\sim 20 \ km$).

\section{\bf Ratio $IU$ over $UGS + ID$ events:}


\begin{figwindow}[1,r,%
{\mbox{\epsfig{file=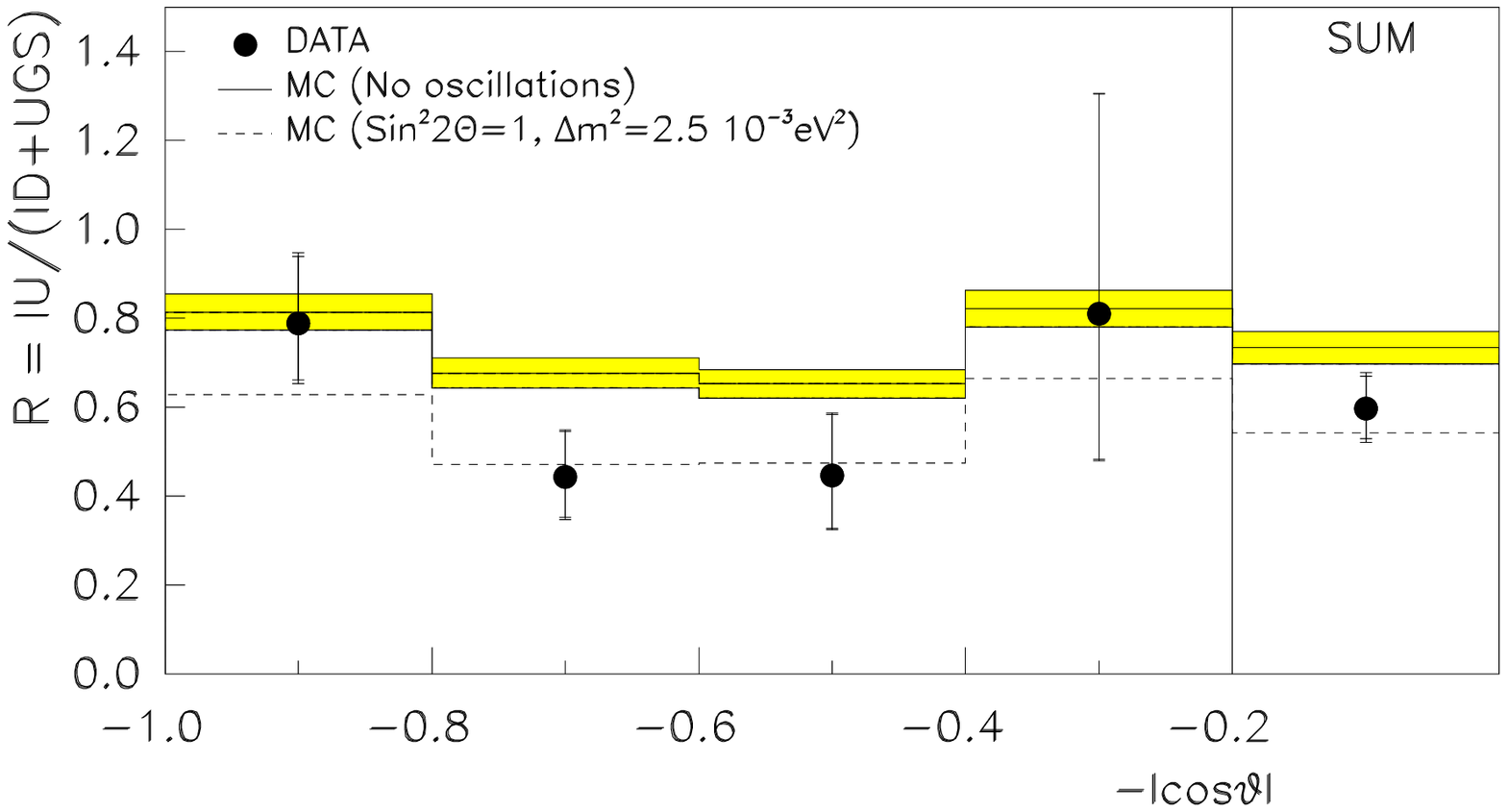,width=3.2in}}},%
{The $cos \theta$ distribution of the ratio between $IU$ and $UGS + ID$ events.
The data result is compared with Monte Carlo expectation assuming no 
oscillation (full line) and two-flavour oscillation (dashed line) using 
maximum mixing and $\Delta m^2=2.5 \times 10^{-3}\ eV^2$.}]
Due to the large theoretical error arising from the uncertainties on absolute 
$\nu$ flux and cross section, the total number of events has a non
negligible probability to be compatible with the no-oscillation hypothesis 
($\sim 6.5 \%$ for $IU$ and $\sim 14 \%$ for $UGS + ID$ events).
On the other side, using the ratio between $IU$ and $UGS + ID$ events,
the theoretical error coming from neutrino flux and cross section 
uncertainties almost disappears. A residual $5 \%$ due to small differences 
between the energy spectra of the two samples survives.
The systematic uncertainty on the ratio is also reduced to $\sim 6 \%$ due 
to some cancellations.
The value of that ratio over the zenith angle distribution obtained from 
data is shown in Fig. 4, where it is compared with MC expectation.
The ratio between the total numbers of detected events is 
$R = 0.60 \pm 0.07_{stat}$, while $R = 0.74 \pm 0.05_{syst} \pm 0.04_{theor}$ 
is expected in case of no oscillation.
The probability to obtain a ratio at least so far from the expected 
one is $\sim 6 \%$ assuming Bartol as the true parent $\nu$ flux
and taking into account the not Gaussian shape of the uncertainty on the
ratio.
In conclusion, the analysis of low energy $\nu$ events collected by 
MACRO shows a preference toward an oscillation model with parameters 
compatible with those suggested by the upward-throughgoing muon data.
\end{figwindow}


%
%
%
\vspace{1ex}
\begin{center}
{\Large\bf References}
\end{center}
%
%
Agrawal, V., et al. 1996, Phys. Rev. D53, 1314\\
Ahlen, S., et al. (MACRO Collaboration) 1993, Nucl. Instr. Meth. A324, 337\\
Allison, W.W.M., et al. 1997, Phys. Lett. B391, 491\\
Ambrosio, M., at al. (MACRO Collaboration) 1998a, Phys. Lett. B434, 451\\
Ambrosio, M., et al. (MACRO Collaboration) 1998b, Astrop. Phys. 9, 105\\
Bernardini, P., (MACRO Collaboration) 1998, Proceedings of "Frontier 
Objects in Astrophysics and Particle Physics - Vulcano Workshop", 
hep-ex/9809003 \\
Casper, D., et al. 1991, Phys. Rev. Lett. 66, 2561\\
%
%
Fukuda, Y., et al. 1994, Phys. Lett. B335, 237\\
Fukuda, Y., et al. 1998, Phys. Rev. Lett. 81, 1562\\
Lipari, P., Lusignoli, M., $\&$ Sartogo, F. 1995, Phys. Rev. Lett. 74, 4384\\
Ronga, F., (MACRO Collaboration) 1999, HE 4.1.07 in this conference\\
Spurio, M., (MACRO Collaboration) 1998, Proceedings 16th ECRS
(Alcala` de Henares, Spain)
\end{document}